# VUV test of a new polarimeter for spectropolarimetric measurements on board space missions

Maëlle Le Gal[a*], Martin Pertenais[b,a], Arturo López Ariste[c], Coralie Neiner[a],
Norbert Champion[d], Youssef Younes[e] and Jean-Michel Reess[a]

[a]LESIA, Observatoire de Paris, PSL Research University, CNRS, Sorbonne Université, Univ. Paris Diderot, Sorbonne Paris Cité, 5 place Jules Janssen, 92195 Meudon, France ; [b]Deutsches Zentrum für Luft- und Raumfahrt e.V., Germany; [c]Institut de Recherche en Astrophysique et Planétologie, CNRS, IRAP, France ; [d]Laboratoire d'Études du Rayonnement et de la Matière en Astrophysique et Atmosphères, Sorbonne Université, Observatoire de Paris, Université PSL, CNRS, LERMA, F-92190, Meudon, France ; [e]GEPI, Sorbonne Université, Observatoire de Paris, Université PSL, CNRS, GEPI, F-92190 , Meudon, France
*maelle.le-gal@obspm.fr

**ABSTRACT**

High-resolution spectropolarimetry is a useful astronomical technique, in particular to study stellar magnetic fields. It has been extensively used in the past but mostly in the visible range. Space missions equipped with high-resolution spectropolarimeters working in the ultra-violet (UV) are now being studied. We propose a concept of a polarimeter working with temporal modulation and allowing to perform Stokes IQUV measurements over the full UV + Visible range. The purpose of this article is to describe the polarimeter concept, two prototypes and the bench developed to perform on ground testing to establish the performances of this new polarimeter.

## 1. INTRODUCTION

Space mission projects including UV high-resolution spectropolarimetry, such as Arago [1] proposed to ESA and Pollux on board LUVOIR [2] proposed to NASA, are being studied in Europe under CNES leadership. Although some instruments have been developed for a use on a small wavelength UV range, spectropolarimeters working on a wide wavelength range including the VUV have never been built and are a true challenge. We propose here a concept of an original polarimeter with temporal modulation allowing to perform simultaneously UV and visible Stokes IQUV measurements on large wavelength range. Two prototypes have been studied, one for Arago that will permit to work from 123 to 888 nm and a second one that has been optimized for the near ultra-violet (NUV) channel of Pollux i.e. from 185 to 400 nm. Having such a large wavelength domain permits the use of the Least Square Deconvolution (LSD) technique [3], i.e. to use many spectral lines simultaneously in order to increase the signal-to-noise ratio of the polarimetric measurement. The simultaneity of the measurements in both UV and visible light is an important aspect because it allows for example to establish a link between what is happening in the stellar environment (such as the wind, chromosphere and magnetosphere, observable in the UV) and what is happening on the stellar surface (such as spots, observable in the visible). Our concept is based on the same principles as a traditional polarimeter: it is basically composed by a light phase modulator followed by an analyzer. Both components are made of magnesium fluoride ($MgF_2$) crystals, the only known birefringent material transparent at wavelengths as short as 123 nm. To demonstrate the feasibility of the instrument, increase the Technical Readiness Level (TRL), and test the performance of our prototype, an optical bench has been set up at the entrance of a vacuum ultra-violet (VUV) spectrometer with a 200 000-spectral resolution at the Meudon Observatory (France). The goal of the test experiment is to be able to put any polarization at the entrance of the bench, first in order to determine the modulation matrix of the prototype and then to determine the feasibility and precision of the measurements. The challenge of such an experiment is the very low global transmission of the bench, as well as the simultaneity of p- and s- polarization measurements with such a high resolution. The purpose of this article is to described both polarimeters, the test experiment and the performances of this instrument.

## 2. PROTOTYPE

### 2.1 Principle

A polarimeter is made of two components, a modulator and an analyzer. For a basic polarimeter, in the visible domain for example, the modulator would be a rotating quarter wave plate and the analyzer would be a basic linear polarizer. For a use in the UV, such simple design cannot be considered. Indeed, at such wavelength, basic materials absorb almost all the flux. In order to answer this need, a new prototype of polarimeter has been developed [4]. The modulator is a rotating modulator made of combined $MgF_2$ plates, with different thicknesses and fast axis angles. A first version was designed for Arago [4], working from 123 to 888 nm, and made of three $MgF_2$ plates. A second version was designed for the NUV channel of Pollux (LUVOIR spectropolarimeter). As shown in figure 1 it is composed of two $MgF_2$ plates and is then thinner, increasing the global transmission. The analyzer is a $MgF_2$ Rochon prism, which has a 100% polarizing efficiency and allow us to retrieve both p and s polarization states, increasing the global efficiency of the polarimeter. These designs have been optimized in order to extract all four Stokes parameters with the best efficiency possible (57.7% for all parameters at the same time). Both polarimeters have been designed to work with 6 measurements, each one corresponds to a rotation of the modulator which we designate as modulation angle. Modulation angles are set for each prototype, they are calculated to have the best efficiency.

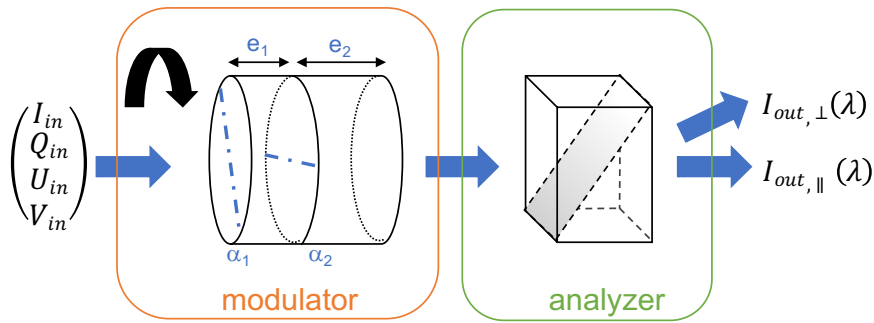

*Figure 1- Polarimeter of Pollux's NUV channel – adapted from [3].*

### 2.2 Modulator characteristics

2.2.1 Modulators

Both modulators have been optimized with six modulation angles and two beams after the analyzer (ordinary and extraordinary beams at the output of the prism), resulting in twelve measurements. A backup version with only six measurements is also considered in case only one beam can be recovered. The modulator of Arago's polarimeter has been designed by Martin Pertenais [4] as a block of three $MgF_2$ waveplates with the following configurations:

- Plate 1: angle of fast axis 0° and thickness 13.5 μm
- Plate 2: angle of fast axis -44° and thickness 8.8 μm
- Plate 3: angle of fast axis 22° and thickness 13.3 μm

This polarimeter has been optimized for a use with 6 modulation angles evenly distributed throughout the field of rotation (180°): 0°, 30°, 60°, 90°, 120° and 150°.

The modulator for Pollux's NUV channel has been optimized as a block of two $MgF_2$ plates with the following configurations:

- Plate 1: angle of fast axis 32.6° and thickness 12.8 μm
- Plate 2: angle of fast axis 147.3° and thickness 3.7 μm

Contrary to the first one, this modulator has been optimized for the following angle of modulation: 1.7°, 11.1°, 34.8°, 50.0°, 69.4° and 111°. This optimization is explained in section 2.2.2.

2.2.2 Optimization

The minimum achievable thickness for such waveplates is 0.3 mm. The study has thus searched in priority to optimize the modulator with feasible plates thicker than 0.3 mm. Unfortunately, no acceptable solutions have been found. In order to realize a modulator with plates of a few microns, each µm-plate is replaced by two feasible mm-plates with a thickness difference of the optimal µm thickness and orthogonal fast axis angle. For example, the first plate of Pollux's NUV modulator will be replaced by two plates, one 0.3 mm plate with a fast axis at 32.6°, and one 0.3128 mm plate with a fast axis at 122.6°.

To choose our final design for Pollux's modulator, we introduce a figure of merit:

$$\kappa = \epsilon \sqrt{T} \qquad (eq.\ 2.2.1)$$

where κ is the figure of merit, ε is the mean of the polarization efficiencies as defined in [5] and T is the global transmission of the modulator. This figure of merit is a criterion of selection considering both efficiency and transmission. The more plates the polarimeter has, the more efficient it is but also the more absorbing it is.

$MgF_2$ complex index of refraction is not well known in this wavelength range. To select our prototype, we studied it at 190 nm, the minimal wavelength with the most absorption. At this wavelength, the linear absorption is $\tau = 0.0083\ mm^{-1}$ [6]. Considering the uncertainties, the calculated figures of merit are the same for the 2 (considering a thickness of 2*0.3 mm for each plate) or 3 plates (about 0.55) if the plates are fixed together with optical contact (molecular adhesion). In fact, most of the loss is due to R, the Fresnel reflection coefficient. This small transmission difference is compensated by the efficiency as one can see in table 1.

$$T = (1-R)^2 e^{-\tau x} \qquad (eq.\ 2.2.2)$$

|  | $MgF_2$ | | | | $SiO_2$ | |
|---|---|---|---|---|---|---|
|  | 2 plates | 2 plates with air gaps | 3 plates | 3 plates with air gaps | 3 plates | 3 plates with air gaps |
| Transmission | 0.929 | 0.767 | 0.924 | 0.672 | 0.87 | 0.46 |
| Efficiency | 0.572 | | 0.577 | | 0.574 | |
| Figure of merit | 0.551 | 0.501 | 0.555 | 0.473 | 0.53 | 0.39 |

*Table 1 - Performances of two prototypes for Pollux's NUV channel at 190nm*

However, if we introduce air gaps, we increase the Fresnel reflection at each dioptre and decrease the transmission. If we introduce air gaps, transmission decreases from 92% to 67% for 3 plates and 76% for a 2-plate modulator. Regarding the performances, it would be better to avoid air gaps, though it may not be possible as it makes the plates more fragile at take-off or in space environment. Indeed, each plate having its own fast axis angle and $MgF_2$ being anisotropic, the stack of plates is very sensitive to thermal expansion. Moreover, the air gaps will introduce fringes in the global transmission which are not considered at this level of the simulation ([7]). Taking that into account, even though the figure of merit is slightly better for a 3-plate polarimeter with optical contact, it seems more convenient to choose a 2-plate polarimeter which has almost the same performances and a slightly better transmission both with or without air gaps. The chosen final prototype's modulator is a rotating 2-$MgF_2$-plate block with the characteristics presented in section 2.2.1 but with each thin plate replaced by a pair of thicker one.

2.2.3 Backup design using $SiO_2$

Using quartz $SiO_2$ has been considered for Pollux's NUV channel as it transmits until 185 nm. Its transmission is slightly below $MgF_2$'s transmission, therefore $MgF_2$ was our first choice. Due to its fragility, as mentioned above, a backup design has been studied using $SiO_2$. As for $MgF_2$, no solution has been found using plates thicker than 0.3 mm. As explained in

2.2.2, we found a solution with μm plates which will define the thickness difference we need between two plates. The study did not result in a 2-plate polarimeter design. The only fulfilling result is a block of 3 μm-plates that need to be doubled. Its characteristics are:

- Plate 1: angle of fast axis 20.1° and thickness 13.1 μm
- Plate 2: angle of fast axis 62.1° and thickness 8.3 μm
- Plate 3: angle of fast axis 168.7° and thickness 12.9 μm

At 190 nm, this modulator has a transmission of 87% for stuck plates and 46% with air gaps and an efficiency of 0.574. The figure of merit is then 0.53 for stuck plates and 0.39 with air gaps.

## 2.3 Theoretical performances

We present here some performances of the $MgF_2$ modulator designed above. To measure the efficiency of the polarimeters, first modulation and demodulation matrices have to be calculated as defined in [5]. Arago's modulation and demodulation matrices are presented in [8]. Modulation and demodulation matrices of Pollux's NUV channel are presented below in figure 2 and figure 3, respectively.

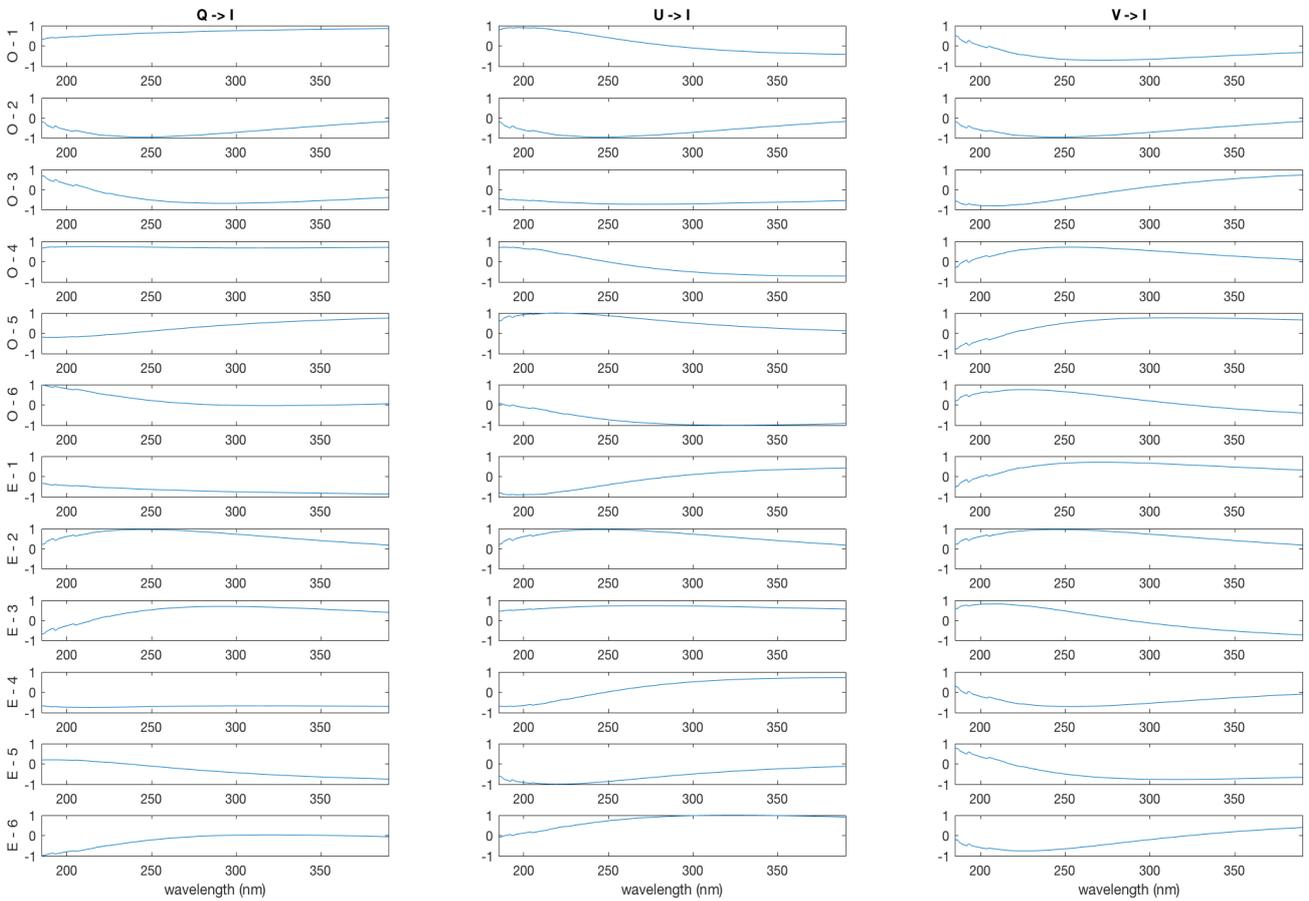

*Figure 2 Modulation matrix – The first column is not represented, it is equal to 1 at all wavelengths. O corresponds to the ordinary beam and E corresponds to the extraordinary beam. The number next to the letter corresponds to the number of the modulation angles.*

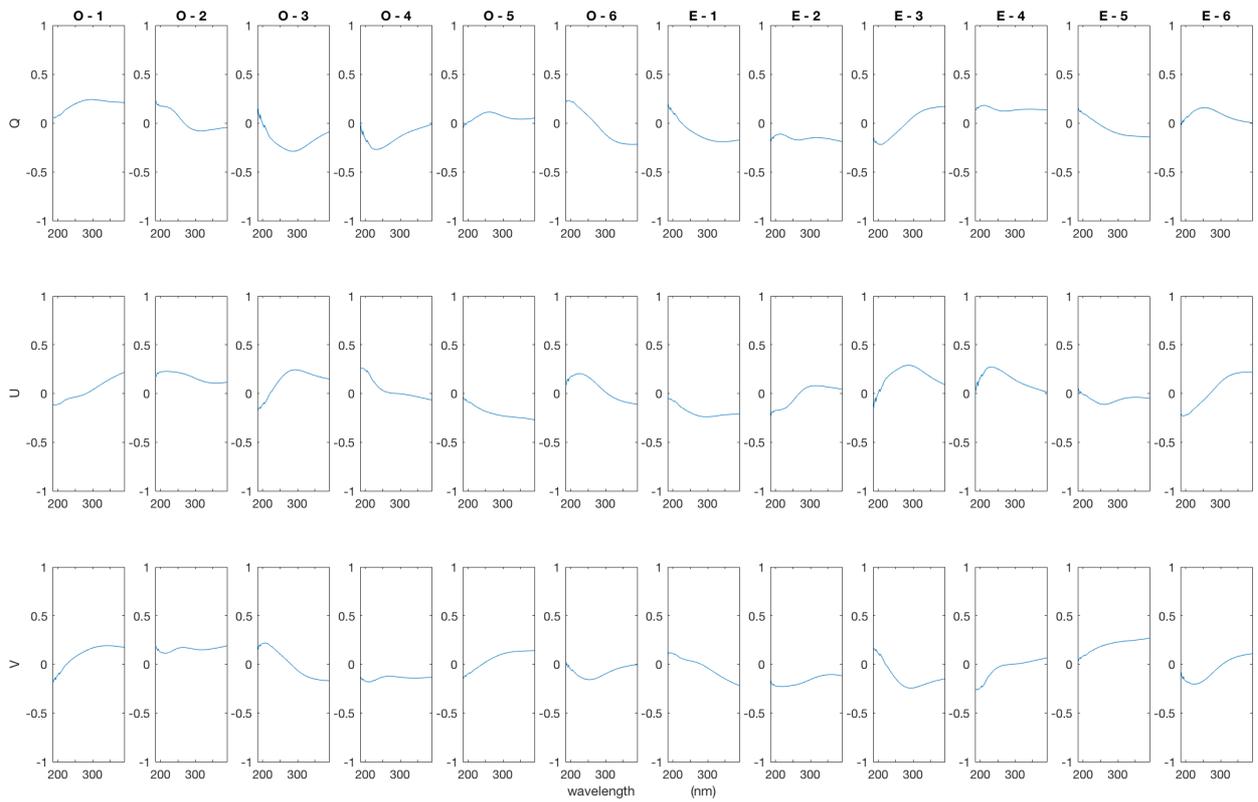

*Figure 3 - Demodulation matrix – The first line is not represented, it is equal to 1 at all wavelengths. O corresponds to the ordinary beam and E corresponds to the extraordinary beam. The number next to the letter corresponds to the number of the modulation angle.*

Polarization efficiencies are presented in figure 4 (left panel). Optimal efficiency being 57.7%, our results are extremely satisfying. The study and the performances are simulated taking into account that both ordinary and extraordinary beams will be recovered at the output of the polarimeter. In case only the ordinary beam can be measured, the polarimeter will still be able to have good results. Indeed, its new polarization efficiencies are presented in figure 4 (right panel) and the mean of polarization efficiencies is now 53%. This result is also satisfying and constitute of a suitable backup solution.

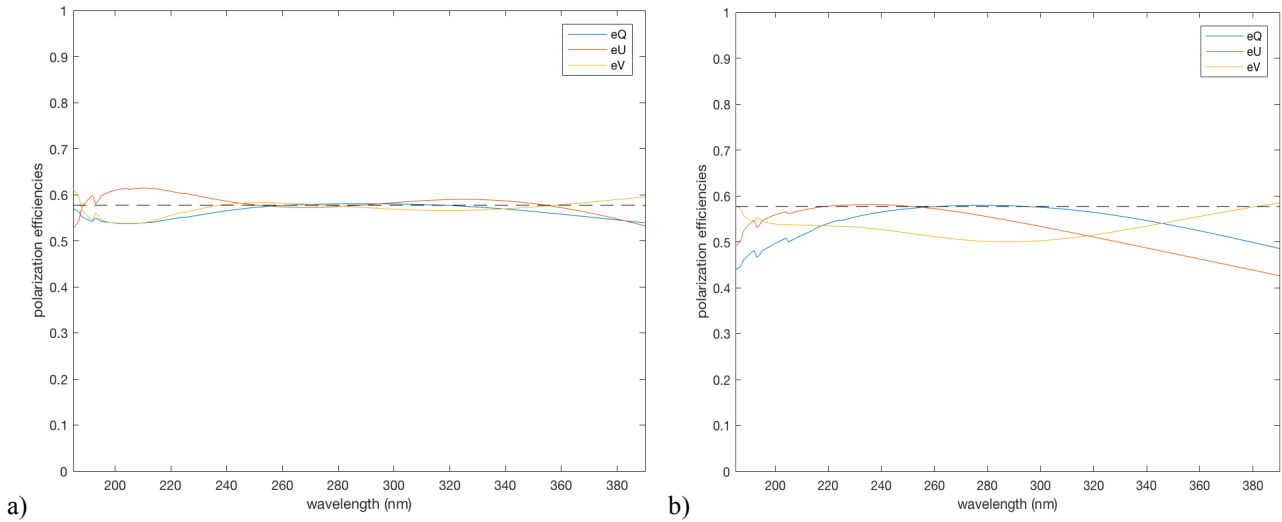

*Figure 4 - Polarization efficiencies as a function of wavelength. Figure (a) is the results for the 2- plate modulator with ordinary and extraordinary beams and figure (b) is for the case where we can only retrieve the ordinary beam.*

In figure 5, one can see the transmission between 180 and 240 nm for the modulator without air gaps (a) and with air gaps (b). In both case, transmission is pretty much constant, hence, the study made at 190 nm can be interpolated over the full wavelength range.

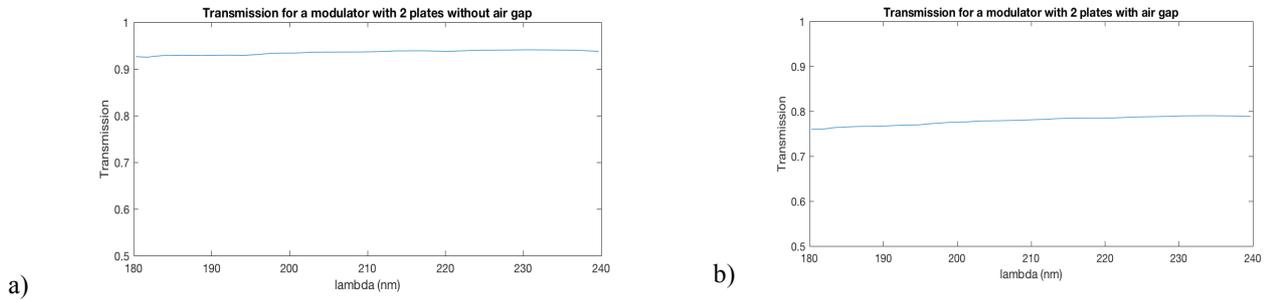

*Figure 5 - Transmission of Pollux NUV channel modulator as a function of wavelength using data from [6].
Figure a) represents the transmission for a modulator with stuck plates and
figure b) represents the transmission for a modulator with air gaps between each plate.*

## 3. TEST BENCH

### 3.1 Principle

The goal of the experiment is to test, validate and establish the performances of the prototypes for a use in the VUV domain. The tests consist of three different measurements. First, the global transmission of the polarimeter has to be measured as a function of wavelength. Then the modulation matrix has to be determined in order to have an experimental demodulation matrix. This is essential, and permits to include all approximations and experimental errors in this matrix, for example polarization due to the reflection on the mirror, spectral calibration or angles uncertainties. This way, the precision will be better. Finally, polarization measurements can be done to determine the performance of the polarimeter in terms of precision and resolution.

Arago's polarimeter has already been tested and validated for visible spectropolarimetric measurements. The purpose of this experiment is to validate VUV spectropolarimetric measurements with this new type of polarimeter.

### 3.2 Optical design

#### 3.2.1 Spectrometer

For the VUV test, a high-resolution spectrometer at LERMA (Observatoire de Meudon, France) has been put at our disposal, permitting to perform high-resolution spectropolarimetric measurements. The spectrometer, represented in figure 6, is a Rowland circle geometry based spectrometer which means the grating, the slit and the detector are on the same virtual circle. Both the grating and the detector are curved. The high resolution is due to the concave grating. Indeed, the grating has 3600 grooves/mm and a curvature of 10.685 m - same as the diameter of the Rowland circle. This grating induces a spectral resolution up to 200 000. The grating's diameter is 23 cm, and the detector is 80 cm long. The spectrometer is at F/23 which is determinant for our optical system. This spectrometer is similar to the one presented in [9].

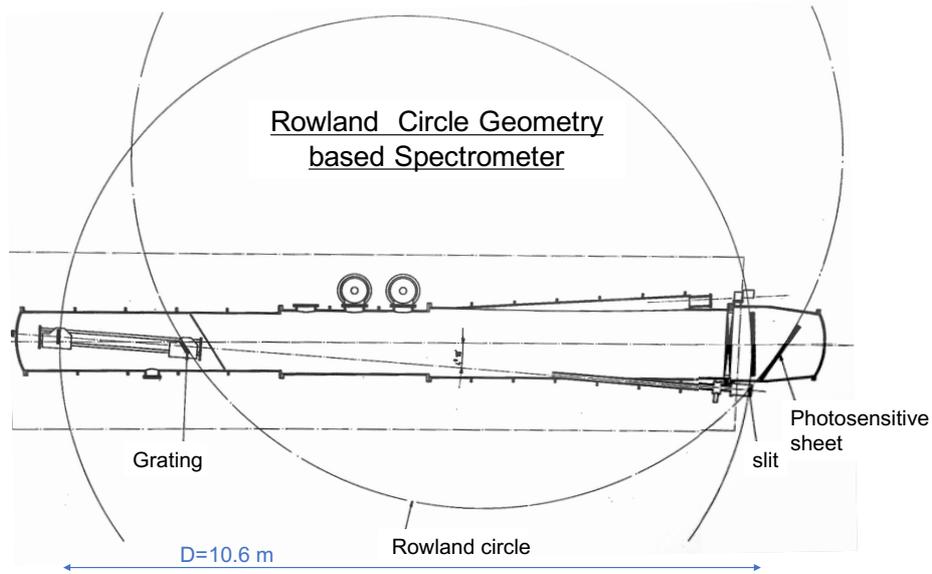

*Figure 6 - Rowland circle geometry based spectrometer.*

### 3.2.2 Optical bench

To test the polarimeters in the UV, a known UV polarized light should be used as an input, and then measured. For this purpose, the test bench should have a UV source, a quarter waveplate and a polarizer in order to create any polarization, followed by the tested polarimeter. As mentioned before, as linear polarizers do not exist at such wavelength, a Rochon prism is used as a polarizer. This choice has as consequences a decreasing of the transmission, first because of its thickness and then because it divides the beam in two. Nevertheless, the polarization has a 100% efficiency unlike Brewster angle polarizer for example. Moreover, a Rochon prism is used given that they are easy to manufacture in $MgF_2$ and allow an easy selection of the polarization - the one which is not deviated – making the alignment easier and avoiding chromatic effects. As a consequence, a diaphragm must be added to completely remove the deviated beam.

In order to make a proper selection of the wanted polarized beam at the output of the Rochon prism, it is convenient to have a collimated beam. Lenses are not a choice here, because of the low transmission and the chromatic effects. Accordingly, we opted for two off-axis parabolic mirrors, which permit to obtain a collimated beam of the source and to project the image of the beam on the entrance slit. These mirrors have been made at *eSource Optics* with a special VUV coating to have a good reflection (>60%) in the studied wavelength range. The final test bench is presented in figure 7. The optical path and the choice of components have been constrained by the mechanics of the spectrometer and the source.

### 3.3 Mechanical design

The mechanical design of this test bench was a challenging part of the set up. The mechanical design had to envelop the optical design as well as to adapt to the already existing mechanical design of the spectrometer. The enclosure has to be vacuum compatible, the polarimeter has to be removable without opening the enclosure, and moreover, the modulator has to rotate automatically. This enclosure is displayed figure 8.

The second parabolic mirror is mounted on a tilt platform which allows to select the measured beam at the output of the polarimeter's Rochon prism. Indeed, both beams could not be measured at the same time due to the spectrometer mechanics.

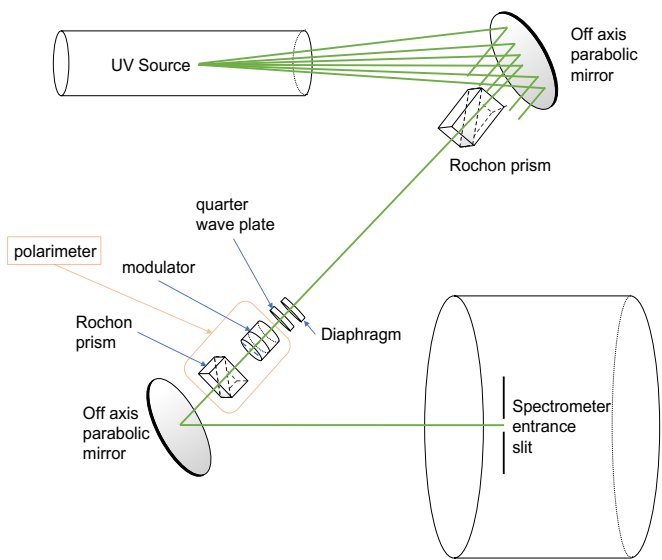 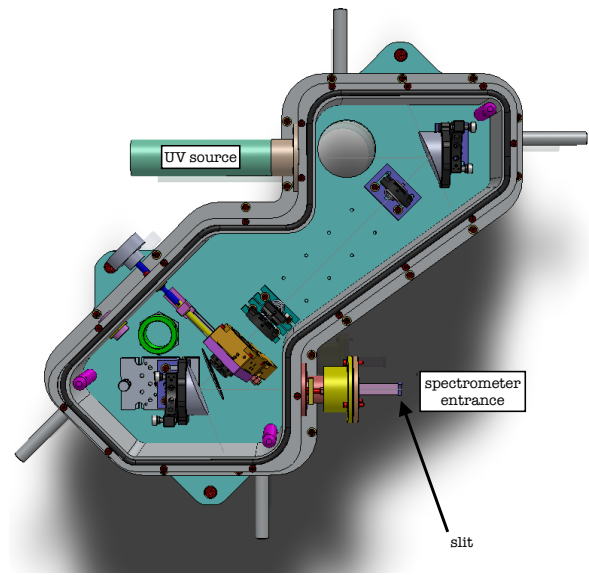

*Figure 7 - Optical design of the VUV test bench - the figure is realistic but not to scale.*   *Figure 8 - Mechanical design of the VUV test bench viewed from the top.*

### 3.4 Performances expectation

As the transmission of $MgF_2$ is not well known, the global transmission of the bench can only be calculated at some precise wavelengths. These values are given in table 2. As one can see, the transmission of the optical bench is quite low as each component have a significant absorption at low wavelengths. This transmission does not take into account the transmission of the polarimeter, which depends on the tested prototype.

To improve the transmission, long exposures are considered to have a significant signal for the test measurements.

|  | @120nm | @130nm | @180nm |
|---|---|---|---|
| Parabolic mirror 1 | 0,77 | 0,77 | 0,73 |
| Rochon prism | 0,0338 | 0,093 | 0,386 |
| Quarter wave plate | 0,68 | 0,766 | 0,915 |
| Parabolic mirror 2 | 0,77 | 0,77 | 0,73 |
| **Total** | **0,0136** | **0,0422** | **0,188** |

*Table 2 - Optical transmission of elements of the VUV test bench.*

Furthermore, part of the flux is lost at the prism's entrance. Indeed, the beam's diameter is larger than the prism's diameter. Unfortunately, this could not be avoided. In order to have a significant transmission of the prism, it was necessary to build a thin one, and thus a small one. Using a mirror with a smaller focal length would have been a great alternative but was not possible in our case due to the mechanical constraints of the UV source and the spectrometer. With the designed polarimeter and this test bench at hand, we will soon start to perform measurements. This will allow us to demonstrate the feasibility of high-resolution spectropolarimetry over a wide wavelength range in the VUV.

## 4. CONCLUSION

Working in the UV domain is a real challenge as very few materials are transparent in the UV. The NUV channel modulator of Pollux has been designed as a rotating block of two $MgF_2$ plates. A backup version has been designed using $SiO_2$. A test bench in a vacuum enclosure has been set up for VUV spectropolarimetric measurements. In the near future, Pollux's NUV polarimeter and Arago's polarimeter are going to be tested and evaluated. At the same time, polarimeters for the MUV and FUV channels of Pollux are being developed as well.

## AKNOWLEDGMENTS

The authors would like to thank CNES for their financial support.
We would like to thank Vartan Arslanyan, Claude Collin and Napoléon Nguyen Tuong for the work they have done for the mechanical pieces, Frank Brachet, Louise Lopes, Josiane Costeraste and Christian Buil for their critical assessments and constructive discussions, and Lydia Tchang-Brillet for putting the spectrometer at our disposal and sharing her knowledge of this instrument.